\documentclass[a4paper,11pt]{article}

\usepackage[english]{babel}
\usepackage[ansinew]{inputenc}
\usepackage{fontenc}
\usepackage{amsfonts}
\usepackage{amsmath}
\usepackage{amssymb}
\usepackage{amsthm}
\usepackage{enumerate}
\usepackage{textcomp}
\usepackage{geometry}
\usepackage{mathrsfs}
\usepackage{natbib}
\usepackage{dsfont} % For the indicator function definition
\usepackage{graphicx}
\usepackage{multirow}
\usepackage{subcaption}
\usepackage{algorithmic}
\usepackage{algorithm}
\usepackage{xcolor}
\numberwithin{equation}{section}

\newcommand{\ind}[1]{\ensuremath{\mathds{1}_{#1}}}

\newtheorem{definition}{Definition}[section]

\newtheorem{proposition}{Proposition}[section]

\theoremstyle{definition}

\begin{document}

\author{Diaa Al Mohamad\footnote{Diaa Al Mohamad is a postdoc researcher at Leiden University Medical Center (LUMC), Einthovenweg 20, 2333 ZC Leiden, The Netherlands (email: diaa.almohamad@gmail.com). This research is funded by the NWO VIDI grant 639.072.412.}\;\;\;\;\; Jelle J. Goeman\footnote{Jelle J. Goeman is a professor in Biostatistics at the LUMC (email: j.j.goeman@lumc.nl)}. \;\; \; \;\; Erik W. van Zwet\footnote{Erik W. van Zwet is an associate professor in medical statistics at the LUMC (email: e.w.van\_zwet@lumc.nl)}  
 \\ 
		\normalsize{Department of Biomedical Data Sciences},
		\\
		\normalsize{Leiden University Medical Center, The Netherlands}
\\
Last update 
}

\title{Simultaneous Confidence Intervals for Ranks With Application to Ranking Institutions}
\maketitle
\begin{abstract}
When a ranking of institutions such as medical centers or universities is based on an indicator provided with a standard error, confidence intervals should be calculated to assess the quality of these ranks. We consider the problem of constructing simultaneous confidence intervals for the ranks of means based on an observed sample. For this aim, the only available method from the literature uses Monte-Carlo simulations and is highly anticonservative especially when the means are close to each other or have ties. We present a novel method based on Tukey's honest significant difference test (HSD). Our new method is on the contrary conservative when there are no ties. By properly rescaling these two methods to the nominal confidence level, they surprisingly perform very similarly. The Monte-Carlo method is however unscalable when the number of institutions is large than 30 to 50 and stays thus anticonservative. We provide extensive simulations to support our claims and the two methods are compared in terms of their simultaneous coverage and their efficiency. We provide a data analysis for 64 hospitals in the Netherlands and compare both methods. Software for our new methods is available online in package \texttt{ICRanks} downloadable from CRAN. Supplementary materials include supplementary R code for the simulations and proofs of the propositions presented in this paper.
\end{abstract}

\section{Introduction}
Estimation of ranks is an important statistical problem which appears in many applications in healthcare, education and social services \citep{GoldsteinSpiegel} to compare the performance of medical centers, universities or more generally institutions. Estimates of ranks have generally a great uncertainty so that confidence intervals (CIs) become crucial \citep{Spiegelhalter,GoldsteinSpiegel}. It is surprising that inference of ranks has received little attention in the statistical literature. In applications, ranks are rarely accompanied with CIs and if so these are generally pointwise. This paper presents a method to produce simultaneous CIs at a prespecified joint level $1-\alpha$ for the ranks with correct coverage of the true ranks. Simultaneity is important in the context of ranking estimation whenever we are not interested in a specific named institution but rather in all the institutions together. Simultaneity is also necessary to quantify the uncertainty about which institutions are ranked best, second best, etc.

In the literature, authors focus usually on pointwise CIs for the ranks. We mention the bootstrap method of \cite{GoldsteinSpiegel} which was widely used, see \citet{Spiegelhalter}, \citet{Gerzoff} and \citet{Feudtner} among others. Methods based on empirical Bayes approaches were also considered, see \citet{LairdThomasBayes}, \citet{Hans}, \citet{LinThomasBayes}, \citet{LinThomasBayesBis}, \citet{LingsmaER}, \citet{NomaBayes} and \cite{Gangnon} among others. We also mention funnel plots, see \citet{TekkisFunnelPlot}, \citet{SpiegelhalterFunnelPlots} among others. These latter two approaches although have been considered in comparing institutions, they do not aim to build (simultaneous) CIs for ranks. On the other hand, it was pointed out by \citet{HallMillerInconsistRank} and \citet{XieMiddleRank} that the bootstrap pointwise CIs and Bayesian methods fail to cover the true ranks in the presence of ties or near ties among the compared institutions. Testing pairwise differences between means was also used to produce pointwise CIs for ranks \citep{LemmersZscore,LemmersZscoreAgain,HolmUnpublished,TingBieMasterThesis}. \citet{LemmersZscore} tested pairwise differences among Dutch hospitals by calculating Z-scores for their performance indicators, but they did not correct for multiple testing and thus their CIs for ranks are not simultaneous. \citet{HolmUnpublished} (see also \citet{TingBieMasterThesis}) calculated also a Z-score, but he applied Holm's sequential algorithm to correct for multiple comparisons on the institution level, that is for each institution he corrects for comparisons with other institutions. Nevertheless, this is only sufficient if we are interested in one of the institutions, but it is not sufficient to produce simultaneous confidence intervals for the ranks of the institutions. 

To the best of our knowledge, there is only one method in the literature introduced by \citet{Zhang} where a Monte-Carlo method is proposed in order to produce simultaneous CIs for ranks. The method was adopted later in some recent papers such as \citet{Waldrop}, \cite{Moss}, \cite{Huang2018} and \cite{Moss2018} among others. The method of \cite{Zhang} can be seen as a generalization of the method proposed by \citet{GoldsteinSpiegel} and can be seen as a parametric bootstrap method. It is however not clear why this method should actually have a simultaneous coverage of at least $1-\alpha$. Besides, since it depends on the method of \citet{GoldsteinSpiegel}, we argue that it inherits the lack of correct coverage. We show through extensive simulations that the method of \cite{Zhang} has the desired simultaneous coverage only when the means are quite far from each other with no way of determining the range of means since this depends on the number of means, how they are scattered and also the standard deviations. We also show that it is anticonservative when the means are close to each other. 

We present a novel method which uses Tukey's honest significant difference (HSD) test \citep{Tukey}. We show that Tukey's HSD can be used to produce simultaneous confidence intervals for ranks with simultaneous coverage of at least $1-\alpha$. 

When the means have no ties, our method becomes conservative. We show that it is possible to adjust the confidence level so that we reduce the conservativeness of the method. We show similarly, it is possible to repair the method of \cite{Zhang} in order to regain control of the confidence level. After rescaling, both methods seem to produce similar results, but the method of \cite{Zhang} becomes extremely difficult to repair as the number of means exceeds 30 to 50.

We also introduce in this paper a new rankability measure defined as the proportion of pairs of institutions that have different true performances. We estimate the true rankability by our method, and provide a lower confidence bound for it. 

The paper is organized as follows. In Section \ref{sec:context}, we explain the context of this paper, the notations and the objective. In Section \ref{subsec:TukeyHSD}, we revisit Tukey's HSD and show that it can be used to provide simultaneous confidence intervals for the ranks. In Section \ref{subsec:Coverage}, we review the Monte-Carlo method of \cite{Zhang}. In Section \ref{sec:RescaledTukey}, we show how to rescale the confidence level of our method and \cite{Zhang}'s method. Our new rankability measure is presented in Section \ref{sec:rankability}. Section \ref{sec:simulations} is devoted to simulation studies comparing our method with \cite{Zhang}'s method with and without rescaling the coverage. An example of ranking Dutch hospitals is also discussed. Proofs of the propositions are in Appendix. Software for the methods presented in this paper is available in package \texttt{ICRanks} downloadable from CRAN.

% =================================================================
% -----------------------------------------
%%%%%%%%%%%%%%%%%%%%%%%%%%%%%%%%%%%%%%%%
% -----------------------------------------
% =================================================================
%%%%%%%%%%%%%%%%%%%%%%%%%%%%%%%%%%%%%%%%%%%%%%%%%%%%%%
\section{Context and Objective}\label{sec:context}
Let $\mu_1,\cdots,\mu_n$ be $n$ real valued numbers which represent for example the true performance of the institutions we want to rank. Let $y=(y_1,\cdots,y_n)$ be a sample of $n$ independent random variables drawn from Gaussian distributions in the following manner
\begin{equation}
y_i \sim \mathcal{N}(\mu_i,\sigma_i^2), \quad \text{for } i\in\{1,\cdots,n\},
\label{eqn:TheGaussModel}
\end{equation}
where the standard deviations $\sigma_1,\cdots,\sigma_n$ are known whereas the centers $\mu_1,\cdots,\mu_n$ are unknown. The sample represents the observed performance indicators. Denote $r_1,\cdots,r_n$ the true ranks of the centers respectively which are the target of inference. Our objective is to build simultaneous CIs for these ranks. Let us first define the ranks $r_1,\cdots,r_n$, allowing for the possibility of ties.
\begin{definition}[ranks]\label{def:ranks}
We define the lower-rank of center $\mu_i$ by 
\begin{equation}
l_i = 1+\sum_{j\neq i}{\ind{\mu_j<\mu_i}}.
\label{eqn:LowerRank}
\end{equation}
We also define the upper-rank of center $\mu_i$ by
\begin{equation}
 u_i = n-\sum_{j\neq i}{\ind{\mu_j\geq \mu_i}}.
\label{eqn:UpperRank}
\end{equation}
We finally define the set-rank of $\mu_i$ as the set of natural numbers $r_i = \{l_i,l_i+1,\cdots,u_i\}$ denoted here $[l_i,u_i]$.
\end{definition}
\paragraph{Assumption 1.} If the means have no ties, then $r_i = l_i = u_i$. Thus, the set-ranks coincide with usual ranking definition; the ranks are calculated for each mean by counting down how many means are below it. 

When there are ties between the centers, we suppose that each of the tied centers possesses a set of ranks $r_i = [l_i,u_i]$. For example, assume that we only have 3 centers $\mu_1,\mu_2$ and $\mu_3$ such that $\mu_1=\mu_2<\mu_3$. Then, the rank of $\mu_1$ is the set $\{1,2\}$ and the rank of $\mu_2$ is also the set $\{1,2\}$, whereas the rank of $\mu_3$ is the singleton $\{3\}$. The rationale of the definition of the set-ranks is that in case of ties, the ranking is arbitrary, and a small perturbation of the true performance may produce any rank in the set of ranks. We call the ranks induced from the observed sample $y$ the empirical ranks. These ranks might be different from the true ranks of the centers, and since the sample is assumed to have a continuous distribution, the empirical ranks are all singletons.
  
We aim on the basis of the sample $y$ to construct simultaneous confidence intervals for the set-ranks of the centers. In other words, for each $i$ we search for a confidence interval $[L_i,U_i]$ such that:
\begin{equation}
\mathbb{P}\left([l_i,u_i]\subseteq [L_i,U_i],\forall i\in\{1,\cdots,n\}\right)\geq 1-\alpha
\label{eqn:SimultCIs}
\end{equation}
for a prespecified confidence level $1-\alpha$. It is worth noting that the confidence intervals here are confidence intervals in $\mathbb{N}$, the set of natural numbers.

Two different types of statement can be obtained from the simultaneous CIs (\ref{eqn:SimultCIs}). First, for each center what are the possible ranks that it might take (which is our main objective). Second, since the confidence intervals for the ranks are simultaneous, we can deduce confidence sets for the best center(s), second best center(s), etc. These confidence sets have also a joint confidence level of at least $1-\alpha$. Indeed, in order to find the centers that can be the best, it suffices to see who are the centers whose rank CI starts at 1. In the same way, we can look at the centers whose rank CI includes rank 2 to obtain a confidence set of the centers ranked second best and so on.

\section{Simultaneous Confidence Intervals for Ranks Using Tukey's HSD}\label{subsec:TukeyHSD}
Tukey's pairwise comparison procedure \citep{Tukey} best known as the honest significant difference test (HSD) is an easy way to compare means of observations  with (assumed) Gaussian distributions especially in ANOVA models. The interesting point about the procedure is that it provides simultaneous confidence statements about the differences between the means and controls the FWER at level $\alpha$. Moreover, it possesses certain optimality properties. In balanced one-way designs (which corresponds in our context to the situation that all $\sigma_i$'s are equal), simultaneous confidence intervals for the differences have confidence level exactly $1-\alpha$. The method is also optimal in the sense that it produces the shortest confidence intervals for all pairwise differences among all procedures that give equal-width confidence intervals at joint level at least $1-\alpha$, see for example \citet[p. 81]{HochbergBook} and \citet{Rafter}.

We consider the general case with possibly unequal $\sigma_i$'s here. Tukey's HSD tests all null hypotheses $H_{i,j}: \mu_i - \mu_j=0$ at level $\alpha$ using the rejection region
\begin{equation}
\left\{\frac{\left|y_i-y_j\right|}{\sqrt{\sigma_i^2 + \sigma_j^2}}>q_{1-\alpha}\right\}
\label{eqn:RejRegionTukey}
\end{equation}
where $q_{1-\alpha}$ is the quantile of order $1-\alpha$ of the distribution of the Studentized range
\begin{equation}
\max_{i,j=1,\cdots,n}\frac{|\tilde{Y}_i-\tilde{Y}_j|}{\sqrt{\sigma_i^2 + \sigma_j^2}},
\label{eqn:StudentizedRange}
\end{equation}
and $\tilde{Y}_1,\cdots,\tilde{Y}_n$ are independent centered Gaussian random variables with standard deviations $\sigma_1,\cdots,\sigma_n$ respectively.

In practice, a simple way to construct the confidence intervals for the ranks is to start by sorting the observations $y_1<y_2<\cdots<y_n$. In order to calculate the CI for $\mu_i$, it suffices to count down how many centers are not significantly different from it. The lower bound of the rank of $\mu_i$ is thus obtained by counting the number of times the hypothesis $\mu_i = \mu_j$ for $j<i$ is not rejected, say $s_i$, or equivalently the number of times the test statistic is below the Studentized range quantile. We then count down how many times the hypothesis $\mu_i = \mu_k$ for $k>i$ is not rejected, say $t_i$. The confidence interval for the rank of $\mu_i$ is then $[i-s_i, i+t_i]$. 
\begin{proposition}\label{prop:TukeySimultCIs}
Let
\begin{align*}
L_i  = & 1+\#\left\{j:\; y_i - y_j - \sqrt{\sigma_i^2 + \sigma_j^2}q_{1-\alpha}>0\right\} \\
U_i  = & n-\#\left\{j:\; y_i - y_j + \sqrt{\sigma_i^2 + \sigma_j^2}q_{1-\alpha}<0\right\}.
\end{align*}
The intervals $[L_i,U_i]$ for $i=1,\cdots,n$ are $(1-\alpha)$-joint confidence intervals for the ranks of means $\mu_1,\cdots,\mu_n$.
\end{proposition} 
%In other words, calculating the confidence interval for the rank of $\mu_i$ is done by counting down how many centers are significantly inferior than it and how many are significantly superior than it. Centers which are not significantly different from the actual center all share the same ranks.and generally larger than the confidence level of the confidence intervals for the centers. This is clear here, because when we transform the "numeric" CIs into "integer" CIs, there will be a loss of information.The proof is in Appendix \ref{Append:TukeySimultCIs}. 
Suppose we have three institutions $A,B$ and $C$ with centers $\mu_A,\mu_B,\mu_C$ respectively. Assume that we found the following $90\%$ confidence intervals for the differences from Tukey's HSD (rounded to 1 digit)
\begin{eqnarray*}
\mu_A - \mu_B \in [-2,-1] &,& \mu_A-\mu_C \in [-3,-2] \\
\mu_B - \mu_A \in [1,2] &,& \mu_B - \mu_C \in [-1,1] \\
\mu_C - \mu_A \in [2,3] &,& \mu_C - \mu_B \in [-1,1].
\end{eqnarray*}
Then center $A$ gets a confidence interval for its rank $[1,1]$, center $B$ gets a confidence interval for its rank $[2,3]$ and center $C$ gets a confidence interval for its rank $[2,3]$.

We do not have a general idea about an optimal procedure to build simultaneous confidence intervals for the ranks. Still, Tukey's HSD is known to have some optimality in balanced designs (all standard deviations are the same). Indeed, it produces the shortest confidence intervals for the differences among all (one-step) procedures which give equal-width confidence intervals at joint level $1-\alpha$. Since this optimality is related directly to the differences among the means, then it leads to the same optimality concerning ranks. In other words, any other method providing confidence intervals for the ranks based on confidence intervals for the differences with equal-width will produce wider CIs for the ranks than the ones produced by our Tukey-based method.

\begin{proposition}\label{prop:TukeyRankExactCov}
Under the full null, that is when $\mu_1=\cdots=\mu_n$, the simultaneous coverage of the confidence intervals $[L_i,U_i]$ for $i=1\cdots,n$ produced by Tukey's HSD is exactly $1-\alpha$. 
\end{proposition}
This result together with Proposition \ref{prop:TukeySimultCIs} state that our Tukey-based method provides simultaneous CIs for the ranks with exact simultaneous confidence level $1-\alpha$ when all the means are the same. Otherwise, the method is conservative. The following sections will shed light on this conservativeness when Assumption 1 holds, that is when there are no ties among the means, and we are going to propose a practical solution that reduces this conservativeness.

Several step-down improvements on Tukey's HSD have been proposed, see \citep{Rafter} for a review. The most efficient and well-known is the REGWQ. Instead of testing equality of pairs of means, the procedure tests blocks of equality of means. Step-down variants of Tukey's HSD control the FWER at level $\alpha$, but do not provide any directional information about the relative position of the centers (no protection against type III errors), so that no information about the ranks can be derived. THe step-down Tukey has not been proven to protect against type III errors, that is the ranking error, (\citep{Welsch}) although some authors believe that it does. Therefore, we decided not to consider this approach to build CIs for ranks.

% =============================================
% ...........................
%%%%%%%%%%%%%%%%%%%%%%%%%%%%%%%%%%%%%%%%%%%%%%%%%%%%%%%%%%%%5
%%%%%%%%%%%%%%%%%%%%%%%%%%%%%%%%%%%%%%%%%%%%%%%%%%%%%%%%%%%%5
% ...........................
% =============================================
\section{Simultaneous Confidence Intervals for Ranks Using Zhang et al.'s Method}\label{subsec:Coverage}
The method of \citet{Zhang} is the first method (as far as we know) in the literature that was proposed to build simultaneous confidence intervals for ranks and can be seen as a generalization of the method of \citet{Spiegelhalter} in some sense. The method proceeds as follows. They use the method introduced by \citet{Spiegelhalter} to produce pointwise CIs at level $1-\beta$ for the ranks for several values of $\beta$ in the interval $(0,\alpha)$. For this purpose, $K$ $n-$samples are generated. These are then used again to estimate by Monte-Carlo the joint probability that the empirical ranks (the ranks of $y$) are inside these pointwise CIs at levels $1-\beta$. They choose $\beta$ such that the set of pointwise CIs has the smallest estimated joint probability superior to $1-\alpha$ that they include the empirical ranks. According to \citet{Zhang}, as $K$ increases, we should obtain simultaneous CIs with a more accurate confidence level superior than $1-\alpha$. They also provide a lower bound for $K$ and advise the reader to choose a sufficiently larger value. 

This Monte-Carlo-based method does not have a solid theoretical assurance and its true simultaneous coverage was never tested on simulated data. In the next sections, we are going to investigate this with more details. We are going to show that the simultaneous coverage of the resulting CIs reaches the nominal level $1-\alpha$ only when the differences among the means are large enough. This depends not only on the range of values of the means, but also on $n$ and the way the means are dispersed in that range. Otherwise, the method is anticonservative and the simultaneous coverage could reach very low levels beyond the nominal level and the resulting CIs would no longer be reliable. On the other hand, we will propose a method to readjust the simultaneous coverage that works in practice only when the number of means is below 50.

%%%%%%%%%%%%%%%%%%%%%%%%%%%%%%%%%%%%%%%%%%%%%%%%%%%%%%%%%%%%5
%% ---------------------------------------
%%%%%%%%%%%%%%%%%%%%%%%%%%%%%%%%%%%%%%%%%%%%%%%%%%%%%%%%%%%%5
\section{Simultaneous confidence intervals for ranks when ties are not allowed}\label{sec:RescaledTukey}
It might be reasonable in the context of ranking to assume Assumption 1, hence the means $\mu_1,\cdots,\mu_n$ are all different, and that there are in reality no ties. We first start by treating the case when the standard deviations are the same, that is $\sigma_1=\cdots=\sigma_n=\sigma$. We move then to the general case of different standard deviations.
%%%%%%%%%%%%%%%%%%%%%%%%%%%%%%%%%%%%%%%%%%%%%%%%%%%%%%%%%%%%5
\subsection{Worst case configuration}
\begin{figure}[ht]
\centering
\includegraphics[scale=0.65]{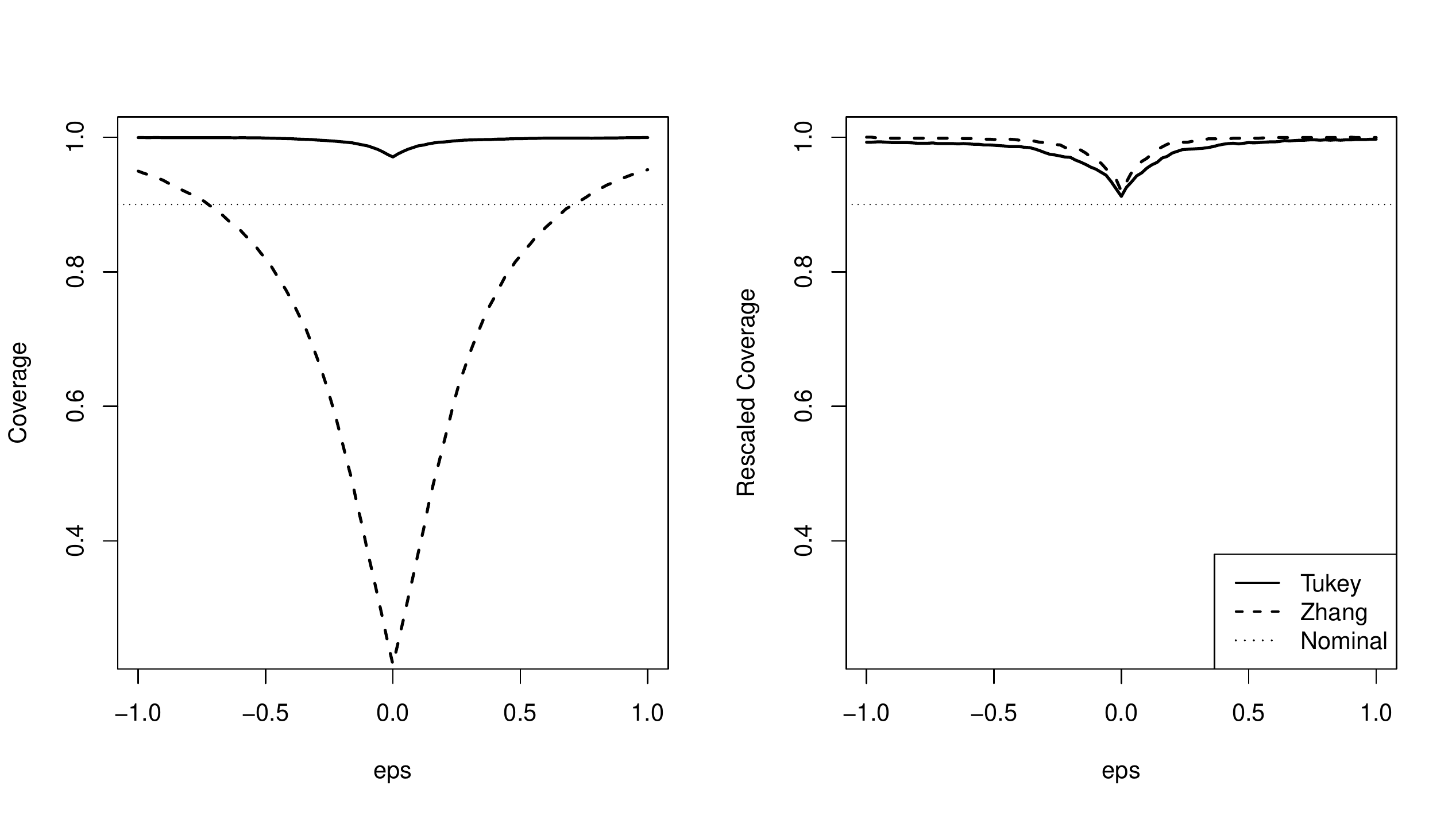}
\caption{The simultaneous coverage of Zhang's'method and our Tukey-based method at vectors of true centers of the form $\varepsilon\mu$ with $\varepsilon\in(-1,1)$. The nominal level is $1-\alpha=0.9$. The left figure corresponds to the actual coverage $\beta_{\mu}(\alpha)$ at joint confidence level $1-\alpha$ whereas the right one corresponds to the actual coverage $\beta_{\mu}(\tilde{\alpha})$ after rescaling the worst case to the nominal level.}
\label{fig:CoverageCurve}
\end{figure}

When there are no ties, our Tukey-based approach becomes more conservative whereas the Monte-Carlo method of \cite{Zhang} becomes less anticonservative. Moreover, as the differences among the means become greater, the coverage probability of these methods should increase. We illustrate this in the left part of figure (\ref{fig:CoverageCurve}) by considering vectors of means of the form $\varepsilon\mu$ where $\varepsilon\in(-1,1)$ and $\mu = (1,\cdots,10)^t$ with dimension 10. The common standard deviation is set to 1. When $\varepsilon=0$, we assume arbitrary ranks for the means, say $1,\cdots,10$, in order to stay conform with the assumption that there are no ties. 

The coverage probability for both methods reaches a minimum when $\varepsilon = 0$. This means that the worst case happens when the means are arbitrarily small, say zero, while not having ties. This means, for any $\mu\neq 0$
\begin{equation}
\mathbb{P}_{\mu}\left(\forall i, r_i\in [L_i,U_i]\right)\geq \mathbb{P}_{\mu=0}\left(\forall i, r_i\in [L_i,U_i]\right).
\label{eqn:WorstCaseCov}
\end{equation}
This worst case configuration is known in hypothesis testing for example when we test if the vector of means has an ascending order (\citet{Robertson78}); the type I error is then highest when all the means are equal. We also mention the Kramer-Tukey procedure (\citet{Tukey,Kramer}); \citet{Hayter} showed that it is conservative and has a worst case configuration when all the standard deviations are the same. In these procedures, the worst case configuration entails exactness of the method, that is the type I error is exactly $\alpha$. In our Tukey-based method, the worst case configuration entails a simultaneous coverage clearly higher than the nominal level. This gap can be further exploited in order to gain more power and reduce the conservativeness of our method. On the other hand, the Monte-Carlo method of \cite{Zhang} can be readjusted so that the simultaneous coverage no longer goes below the nominal level.
%In order to illustrate the worst case for both our Tukey-based method and Zhang's method, we fix $\alpha=0.1$ and consider the vector of true means $\mu=(1,\cdots,10)$ with length 10. We then estimate the coverage for both methods by generating Gaussian random vectors $\mathcal{N}(\varepsilon\mu, I_n)$ for values of $\varepsilon \in (-1,1)$. The number of simulations is $10^5$ for each value of $\varepsilon$, see figure (\ref{fig:CoverageCurve}).
%\begin{table}[ht]
%\centering
%\begin{tabular}{cccccc}
 %& \multicolumn{2}{c}{Coverage at $\alpha=0.05$} & & \multicolumn{2}{c}{Coverage at $\alpha=0.2$} \\
%\cline{2-3} \cline{5-6}
 %& Zhang & Tukey & & Zhang & Tukey \\
%$n=10$ & 0.363 & 0.992 & & 0.130 & 0.930 \\
%$n=30$ & 0.106 & 0.995 && 0.012 & 0.971 \\
%$n=100$ & 0.004 & 0.999 && 0.000 & 0.992  
%\end{tabular}
%\caption{Coverage of Zhang's method and our Tukey-based method at the worst case when ties are not allowed.}
%\label{tab:WorstCaseCoverage}
%\end{table}
%%%%%%%%%%%%%%%%%%%%%%%%%%%%%%%%%%%%%%%%%%%%%%%%
\subsection{Rescaling the coverage at the worst case to the nominal level}
We propose to regularize both methods (ours and \citet{Zhang}'s) so that on the one hand, the Monte-Carlo method of \cite{Zhang} delivers always a simultaneous coverage of at least $1-\alpha$ (a scaling up), and on the other hand, our Tukey-based method delivers a simultaneous coverage of at least $1-\alpha$ but in a less conservative way (a scaling down). Let $\mu$ be some vector of means, assume that at the simultaneous confidence level $1-\alpha$ we get an actual coverage of $\beta_{\mu}(\alpha)$ that is
\[\mathbb{P}_{\mu}\left(\forall i, r_i\in [L_i(\alpha),U_i(\alpha)]\right) = \beta_{\mu}(\alpha)\neq 1-\alpha.\]
We look for $\tilde{\alpha}$ such that
\[\mathbb{P}_{\mu}\left(\forall i, r_i\in [L_i(\tilde{\alpha}),U_i(\tilde{\alpha})]\right) = \beta_{\mu}(\tilde{\alpha})= 1-\alpha.\]
In other words, we look for a zero of the function $z\mapsto \beta_{\mu}(z) - 1+\alpha$ in the interval $(\alpha,1)$ for our Tukey-based method, and in the interval $(0,\alpha)$ for the Monte-Carlo method of \cite{Zhang}. This can be performed using any mathematical program, for example using function \texttt{uniroot} available in the statistical program R (\cite{RProg}). In practice, this is not feasible because $\mu$ is unknown. Therefore, we rescale the coverage at the worst case and use the rescaled confidence level in order to calculate the CIs for any (unknown) $\mu$ different from the null vector. Since the worst-case configuration verifies (\ref{eqn:WorstCaseCov}), the resulting CIs have simultaneous coverage of at least $1-\alpha$. 

The worst case configuration is characterized by having all means infinitely small. Therefore, in practice we set $\mu=(0,\cdots,0)\in\mathbb{R}^n$ and assume arbitrary ranks to its coordinates, for example $1,\cdots,n$. Then we look for $\tilde{\alpha}$ such that
\[\mathbb{P}_{\mu=0}\left(\forall i, r_i\in [L_i(\tilde{\alpha}),U_i(\tilde{\alpha})]\right) = \beta_0(\tilde{\alpha})= 1-\alpha.\]
We have now, for any $\mu\neq 0$,
\[\mathbb{P}_{\mu}\left(\forall i, r_i\in [L_i(\tilde{\alpha}),U_i(\tilde{\alpha})]\right)\geq \mathbb{P}_{\mu=0}\left(\forall i, r_i\in [L_i(\tilde{\alpha}),U_i(\tilde{\alpha})]\right) = 1-\alpha.\]
If we do so, the simultaneous coverage of both our Tukey-based method and the Monte-Carlo method of \cite{Zhang} will be equal to the nominal level $1-\alpha$ near zero and higher than $1-\alpha$ elsewhere as illustrated in the right part of figure (\ref{fig:CoverageCurve}).

Table (\ref{tab:RescaledAlpha}) shows the rescaled significance level $\tilde{\alpha}$ necessary to reach an actual coverage of $80\%, 90\%$ and $95\%$ when the number of centers increases from $10$ to $100$. The table shows that in order to use the method of \cite{Zhang} and make sure not to be anticonservative, we need to use very small values of the significance level. However, as $\tilde{\alpha}$ becomes smaller we need to increase the number of Monte-Carlo samples $K$ required to estimate the joint distribution of the ranks as mentioned by \cite{Zhang}. For example, when $n=50$ it is required that we generate at least $K=10^6$ $n-$samples, and thus rescaling the method of \cite{Zhang} becomes quickly infeasible for higher number of means so that the resulting CIs are not ensured to have the desired coverage of $1-\alpha$. On the other hand, our Tukey-based method although the rescaled significance level moves towards 1, the resulting CIs will always have simultaneous coverage of at least $1-\alpha$ even if we do not fully rescale the significance level at the worst configuration to $1-\alpha$.

\begin{table}[ht]
\centering
\begin{tabular}{ccccccccc}
 & \multicolumn{8}{c}{Rescaled coverage}\\
\cline{3-8}
& \multicolumn{2}{c}{$95\%$} & & \multicolumn{2}{c}{$90\%$}& & \multicolumn{2}{c}{$80\%$} \\
\cline{2-3} \cline{5-6} \cline{8-9}
       & Tukey & Zhang & & Tukey & Zhang & & Tukey & Zhang \\
$n=10$ & 0.158 & $6.5\times 10^{-4}$ & & 0.285 & 0.0015 & & 0.467 & 0.006\\
$n=30$ & 0.303 & $9.8\times 10^{-6}$ & & 0.491 & $4.6\times 10^{-5}$ & & 0.693 & $4\times 10^{-5}$ \\
$n=50$ & 0.418 & $<5\times 10^{-6}$ & & 0.574 & $7\times 10^{-6}$ & & 0.778 & $3.1\times 10^{-5}$ \\
$n=100$ & 0.545 & $<5\times 10^{-6}$ & & 0.725 & $<5\times 10^{-6}$ & & 0.893 & $5\times 10^{-6}$
\end{tabular}
\caption{Values of $\tilde{\alpha}$ necessary to rescale the coverage at worst case back to $1-\alpha$.}
\label{tab:RescaledAlpha}
\end{table}
%
%If we trace the worst case coverage for both methods as a function of the number of means $n$, we realize that starting from $n=30$, it becomes nearly 0 for Zhang's method. This means that it is very difficult to rescale it towards $95\%$ or $90\%$ without using a significance level $\alpha$ extremely small. We are already at $\alpha=10^-5$ for $n=30$. This means that we need to consider a bootstrap sample of order at least $10^5$ in order to perform Zhang's method once and make sure that the simultaneous coverage is at least $90\%$. For our Tukey-based method, the problem is rather reversed. The coverage is almost $100\%$ for $n=100$. This means that we need to use a very high $\alpha$ in order to reduce the conservativeness and get a simultaneous coverage around $90\%$. Still, our method will always cover even without fully rescaling towards the desired $1-\alpha$ whereas for Zhang's method, if we do not rescale correctly, the method is not guaranteed to cover.
%\begin{figure}[ht]
%\centering
%\includegraphics[scale = 0.6]{CoverageAtWorstCase.pdf}
%\caption{Change in the worst case coverage for both Zhang's method and our Tukey-based method as a function of the number of centers $n$.}
%\label{fig:CoverageAtWorstCase}
%\end{figure}

\subsection{The unequal sigma case}\label{subsec:RescalUneqSigma}
When the standard deviations are not the same, the worst case configuration still happens when the means are arbitrarily close to each other, but the order of the $\sigma$'s has an influence on it. Therefore, we need to find a worst-case ordering of the $\sigma$'s which ensures that if we protect against it by properly rescaling the worst case, the method will be conservative for any other ordering. Still, the number of possibilities is huge, that is $n!/2$ possible configurations of the standard deviations. By inspecting these configurations in the case when $n=10$, we found out that the worst-case ordering of the $\sigma$'s should happen when the lowest standard deviations are attributed to the lowest and the largest ranked centers whereas the highest standard deviations are attributed to the middle ranks (for example configuration (\ref{eqn:ConfigTree1})).

\begin{figure}[ht]
\centering
\includegraphics[scale=0.6]{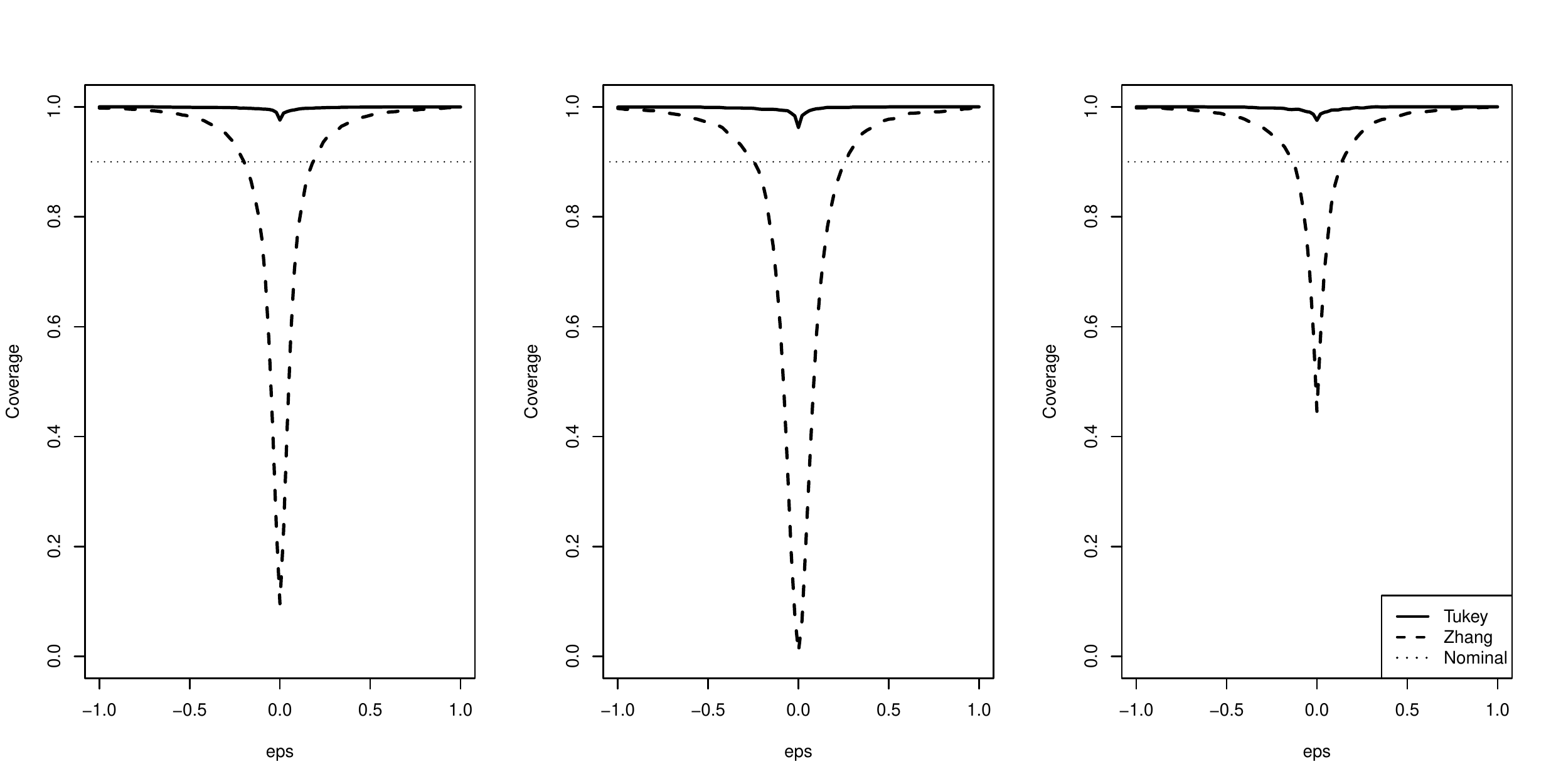}
\caption{Left figure for ascending order (\ref{eqn:ConfigAscend3}), middle figure is for (\ref{eqn:ConfigTree1}) and right figure is for (\ref{eqn:ConfigTree2}).}
\label{fig:CoverageCurveDiffSigmasAll}
\end{figure}

We show in three different scenarios how the worst case scenario happens nearly at zero, figure (\ref{fig:CoverageCurveDiffSigmasAll}). We consider the vector of standard deviations $(1/10,1/9,\cdots,1)$ and a vector of means $\mu=(1,\cdots,10)$. We set $\varepsilon\in (-1,1)$ as before and estimate the coverage for the vectors $\varepsilon\mu$ when the standard deviations are ordered in one of the following manners (two tree orderings and one ascending order)
\begin{align}
\sigma_1\leq \cdots \leq\sigma_n &, \label{eqn:ConfigAscend3} \\
\sigma_1\leq \cdots\leq\sigma_{n/2}, & \sigma_{n/2}\geq \cdots\geq\sigma_n, \label{eqn:ConfigTree1}\\
\sigma_1\geq \cdots\geq\sigma_{n/2}, & \sigma_{n/2}\leq \cdots\leq\sigma_n. \label{eqn:ConfigTree2} 
\end{align}
The estimated coverage at the worst case configuration attains its smallest value under configuration (\ref{eqn:ConfigTree1}).

%(when $\varepsilon=0$) is given in table (\ref{tab:WorstCaseDiffSigDiffScenario}) for $\alpha=0.1$. We calculate as well the rescaled significance level $\tilde{\alpha}$ which corresponds to having actually a coverage equal to the nominal confidence level $1-\alpha=90\%$ at the worst case configuration. The results support our intuition that the worst case configuration happens when the means are all equal and the vector of standard deviations has the tree ordering (\ref{eqn:ConfigTree1}).
%\begin{table}[ht]
%\centering
%\begin{tabular}{cccccc}
 %& \multicolumn{2}{c}{Coverage at $1-\alpha$} && \multicolumn{2}{c}{Rescaled $\alpha$} \\
%\cline{2-3} \cline{5-6}
 %& Tukey & Zhang & & Tukey & Zhang \\
%Config (\ref{eqn:ConfigAscend3}) & 0.976 & 0.096 & & 0.243 & 0.0006? \\
%Config (\ref{eqn:ConfigTree1}) & 0.963 & 0.008 & & 0.182 & 0.0002? \\
%Config (\ref{eqn:ConfigTree2}) & 0.978 & 0.453 & & 0.281 & 0.0091? 
%\end{tabular}
%\caption{Coverage at the worst case scenario for the three different configurations of the standard deviations (\ref{eqn:ConfigAscend3}, \ref{eqn:ConfigTree1},\ref{eqn:ConfigTree2}) when $\alpha=0.1$. The required $\tilde{\alpha}$ for the actual coverage to be $1-\alpha$ is also shown.}
%\label{tab:WorstCaseDiffSigDiffScenario}
%\end{table}
%%%%%%%%%%%%%%%%%%%%%%%%%%%%%%%%%%%%%%%%%%%%%%%%%%%%%%%%%%%%5
%%%%%%%%%%%%%%%%%%%%%%%%%%%%%%%%%%%%%%%%%%%%%%%%%%%%%%%%%%%%5
\section{A Rankability Measure}\label{sec:rankability}
It is useful to have a single measure that gives an impression how well we can distinguish different means, that is how rankable they are. A set of equal centers is evidently not rankable. Therefore, this set of centers should get a rankability of 0. On the other hand, a set of totally different centers should get a rankability of 1 (or $100\%$) since we can rank each center. As the ranks are observed through quantities provided with uncertainty, an estimate of the "true" rankability should be considered along with a confidence interval. We will first define the estimand before we define the estimate and its CI.

\begin{figure}[ht]
\centering
\includegraphics[scale=0.7]{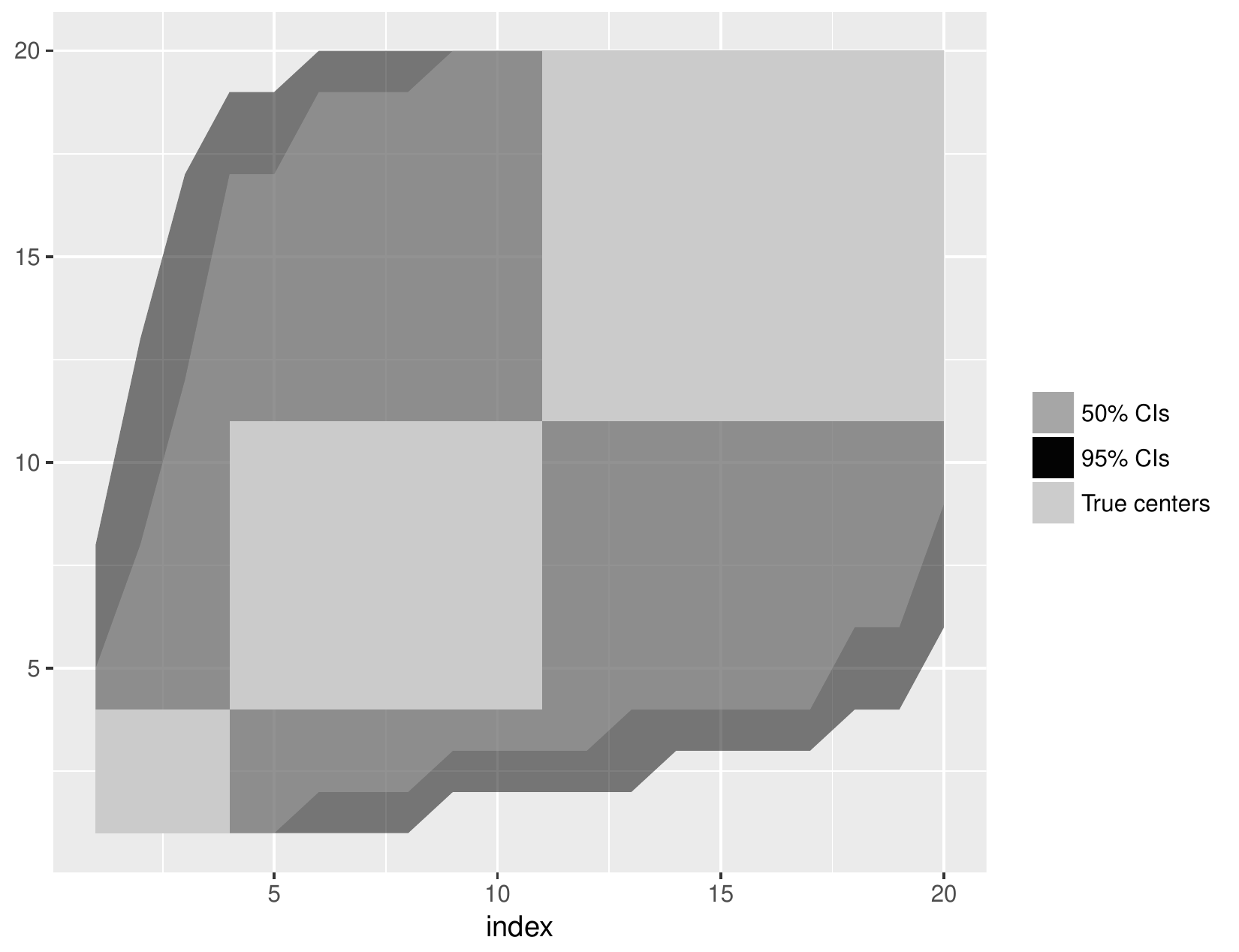}
\caption{Underestimating the Rankability $R_n$. A simulated example showing $95\%$ and $50\%$ simultaneous CIs for the ranks of a set of centers forming three distinct blocks. The (normalized) surface of the light grey blocks is equal to $1-R_n$. The normalized surface of the $50\%$ ($95\%$ resp.) simultaneous CIs gives an underestimation of $R_n$ with a probability $50\%$ ($95\%$ resp.).}
\label{fig:rankability}
\end{figure}

Assume we have $n$ centers $\mu_1\leq\cdots\leq\mu_n$. Some of these centers might be equal. According to our definition of ranks, equal (or tied) centers all get a set of ranks $[l_i,u_i]$ which is the same for all of them. Define the rankability $R_n$ by
\[R_n = 1-\frac{1}{n(n-1)}\sum_{i=1}^n{(u_i-l_i)}.\]
The normalization by $n(n-1)$ is necessary for the rankability $R_n$ to take values in the interval $[0,1]$. The sum gives the surface of the set-ranks (the light grey area in figure (\ref{fig:rankability})) and the subtraction from one ensures that if the set-ranks cover the whole range of ranks, we conclude that the centers are not rankable and we say then that the set of centers have a rankability of 0. In figure (\ref{fig:rankability}), the true rankability is $R_{20} = 0.616$. The surface of the region in light grey (normalized by $n(n-1)$) in figure (\ref{fig:rankability}) can be interpreted as the probability that two centers $\mu_i$ and $\mu_j$ picked at random have the same rank. Therefore, our rankability measure $R_n$ can be interpreted as the probability that two centers picked at random get different ranks.

The rankability $R_n$, since it is defined through the true set-ranks, is a parameter that may be estimated. Denote $[L_i,U_i]$ the confidence interval for the set-rank of $\mu_i$. We assume that these CIs have joint confidence level of $1-\alpha$, that is
\begin{equation}
\mathbb{P}\left(\forall i=1,\cdots,n \;\; [l_i,u_i]\subseteq [L_i,U_i]\right) \geq 1-\alpha.
\label{eqn:SimultCIsSets}
\end{equation}
Define the estimated rankability at level $1-\alpha$ by
\begin{equation}
\hat{R}_n(\alpha) = 1-\frac{1}{n(n-1)}\sum_{i=1}^n{(U_i-L_i)}.
\label{eqn:EstimRankability}
\end{equation}
Due to inequality (\ref{eqn:SimultCIsSets}), the estimated rankability at level $1-\alpha$ underestimates the true rankability with a probability at least $1-\alpha$. In other words
\[\mathbb{P}\left(R_n\geq \hat{R}_n(\alpha)\right) \geq 1-\alpha.\]
Since $R_n\in [0,1]$, the interval $[\hat{R}_n(\alpha),1]$ becomes a $1-\alpha$ confidence interval for $R_n$.

In figure (\ref{fig:rankability}), we show the $50\%$ simultaneous CIs for ranks calculated using Tukey's HSD on a sample of 20 centers resulting in a $50\%$ CI for the $R_n$ which is $[0.232,1]$. $\hat{R}_n(0.5)$ underestimates the true rankability $R_n$ with probability at least $50\%$, and it thus overestimates it with probability at most $50\%$ as well which makes from $\hat{R}_n(0.5)$ a good candidate for a conservative point estimate of $R_n$. We also show in figure (\ref{fig:rankability}) the $95\%$ simultaneous CIs produced by Tukey's HSD, and the resulting $95\%$ CI for $R_n$ is then $[0.126,1]$.
It is worth mentioning that the estimated rankability can also be seen as a performance (or error) index so that several methods providing simultaneous CIs can be compared based on their estimated rankability.  

In the context of empirical Bayesian methods for estimating ranks, a rankability measure was proposed by \citet{Hans}. It indicates which part of variation between hospitals is due to true difference and which part is due to chance. Rankability is then computed by relating heterogeneity between the centers to uncertainty between and within the centers, see also \citet{LingsmaER} and \citet{Fiocco} among others. This measure is specific to the Bayesian method and cannot be used in our case, however our rankability measure is only related to the confidence intervals regardless of the method which produce them. The only requirement is that the confidence intervals are simultaneous.

% ...........................
%%%%%%%%%%%%%%%%%%%%%%%%%%%%%%%%%%%%%%%%%%%%%%%%%%%%%%%%%%%%5
%%%%%%%%%%%%%%%%%%%%%%%%%%%%%%%%%%%%%%%%%%%%%%%%%%%%%%%%%%%%5
% ...........................
% =============================================
\section{Simulation Study and Real Data Analysis}\label{sec:simulations}
In this section we provide several examples (real and simulated) to demonstrate the confidence intervals produced using our approaches from Sections \ref{subsec:TukeyHSD} and \ref{sec:RescaledTukey}. We also compare the coverage and the efficiency of the confidence intervals produced by the method proposed by \citet{Zhang} to the ones obtained by our method in different simulated scenarios. The efficiency is calculate as $1-\hat{R}_n$ where $\hat{R}_n$ is given by (\ref{eqn:EstimRankability}). It represents the average lengths of the CIs.

Finally, we consider a dataset for patients with abdominal aneurysms from 64 hospitals in the Netherlands. We compare these hospitals according to the mortality rate at 30 days and then according to the type of surgery operated on the patient. All the simulations and the data analysis are done using the statistical program \citet{RProg}, and the code of the functions is available in the R package \texttt{ICRanks} which can be downloaded from the CRAN repository. The R code for the method of \cite{Zhang} is provided in the supplementary materials.
%This simulation can illustrate either a context of ranking institutions or experiments in agriculture such as breeding experiments or variety trials. In the later, the number of genotypes to be compared may be in tens or hundreds according to the study, see \citet{CarmerSwanson}, \citet{CarmerWalker} and package \texttt{agricolae} of Felipe de Mendiburu.
% -----------------------------------------
% =================================================================
%%%%%%%%%%%%%%%%%%%%%%%%%%%%%%%%%%%%%%%%%%%%%%%%%%%%%%
%%%%%%%%%%%%%%%%%%%%%%%%%%%%%%%%%%%%%%%%%%%%%%%%%%%%%%%%%%%%%%%%%%
% --------------------------------------
%%%%%%%%%%%%%%%%%%%%%%%%%%%%%%%%%%%%%%%%%%%%%%%%%%%%%%%%%%%%%%%%%%
\subsection{The case of a common standard deviation}\label{subsec:SimCoverage}
The simulation setup is the following. We aim to estimate the average simultaneous coverage of the Monte-Carlo method of \cite{Zhang} and our Tukey-based method. To do so, we generate the centers $\mu_i$'s independently from the Gaussian distribution $\mathcal{N}(0,\tau^2)$ for $\tau = 0.5, 1, 2$. For each value of $\tau$, we generate $1000$ $n-$samples of centers $\mu=(\mu_1,\cdots,\mu_n)$ for $n=10,30$ and $50$. Then a Gaussian vector $y$ is generated from the multivariate Gaussian distribution $\mathcal{N}(\mu,I_n)$. The simultaneous coverage based on these samples is estimated. The rescaled values of $\alpha$ for both methods are already calculated in table (\ref{tab:RescaledAlpha}). We provide a table of the estimated coverage before and after rescaling the significance level so that the actual coverage at the worst case becomes $1-\alpha$ for $\alpha=0.1$. We calculate also the average $1-\hat{R}_n(\alpha)$ where $\hat{R}_n(\alpha)$ is the rankability measure (\ref{eqn:EstimRankability}).
%Standard deviation at n=10 was 0.01, n=30 was 0.005, n=50 was 0.003.
%Standard deviation for n=10 was 0.04, n=30 was 0.005, n=50 is around 0.025.
% Standard deviation of $1-\hat{R}_n(\alpha)$ is around $0.1$.
\begin{table}[ht]
\centering
\begin{tabular}{cccccccccccc}
 & \multicolumn{5}{c}{Coverage} & & \multicolumn{5}{c}{$1-\hat{R}_n(\alpha)$} \\
\cline{2-6} \cline{8-12} \\
 & \multicolumn{2}{c}{not rescaled} & & \multicolumn{2}{c}{rescaled} & &  \multicolumn{2}{c}{not rescaled} & & \multicolumn{2}{c}{rescaled} \\
\cline{2-3} \cline{5-6} \cline{8-9} \cline{11-12}
 & Tukey & Zhang & & Tukey & Zhang & & Tukey & Zhang & & Tukey & Zhang\\
$n=10$ & 0.998 & 0.468 & & 0.961 & 0.976 & & 0.990 & 0.789 & & 0.971 & 0.977 \\  
$n=30$ & 1.000 & 0.027 & & 0.978 & 0.987 & & 0.998 & 0.740 & & 0.990 & 0.991 \\
$n=50$ & 0.997 & 0.000 && 0.976 & 0.984 && 0.999 & 0.726 && 0.994 & 0.995
\end{tabular}
\caption{Coverage probability and the efficiency when $\tau=0.5$ and $\alpha=0.1$ before and after rescaling to the worst case. }
\end{table}

\begin{table}[ht]
\centering
\begin{tabular}{cccccccccccc}
 & \multicolumn{5}{c}{Coverage} & & \multicolumn{5}{c}{$1-\hat{R}_n(\alpha)$} \\
\cline{2-6} \cline{8-12} \\
 & \multicolumn{2}{c}{not rescaled} & & \multicolumn{2}{c}{rescaled} & &  \multicolumn{2}{c}{not rescaled} & & \multicolumn{2}{c}{rescaled} \\
\cline{2-3} \cline{5-6} \cline{8-9} \cline{11-12}
 & Tukey & Zhang & & Tukey & Zhang & & Tukey & Zhang & & Tukey & Zhang\\
$n=10$ & 0.996 & 0.603 & & 0.972 & 0.994 & & 0.959 & 0.698 & & 0.916 & 0.935 \\  
$n=30$ & 1.000 & 0.088 & & 0.993 & 0.996 & & 0.987 & 0.651 & & 0.957 & 0.967 \\
$n=50$ & 0.999 & 0.016 && 0.996 & 0.998 && 0.992 & 0.640 && 0.970 & 0.976
\end{tabular}
\caption{Coverage probability and the rankability when $\tau=1$ and $\alpha=0.1$ before and after rescaling to the worst case. }
\end{table}

\begin{table}[ht]
\centering
\begin{tabular}{cccccccccccc}
 & \multicolumn{5}{c}{Coverage} & & \multicolumn{5}{c}{$1-\hat{R}_n(\alpha)$} \\
\cline{2-6} \cline{8-12} \\
 & \multicolumn{2}{c}{not rescaled} & & \multicolumn{2}{c}{rescaled} & &  \multicolumn{2}{c}{not rescaled} & & \multicolumn{2}{c}{rescaled} \\
\cline{2-3} \cline{5-6} \cline{8-9} \cline{11-12}
 & Tukey & Zhang & & Tukey & Zhang & & Tukey & Zhang & & Tukey & Zhang\\
$n=10$ & 0.997 & 0.814 & & 0.989 & 0.997 & & 0.811 & 0.529 & & 0.734 & 0.788 \\  
$n=30$ & 0.998 & 0.262 & & 0.988 & 0.996 & & 0.888 & 0.479 & & 0.802 & 0.844 \\
$n=50$ & 1.000 & 0.065 && 0.997 & 1.000 && 0.911 & 0.468 && 0.831 & 0.867
\end{tabular}
\caption{Coverage probability and the rankability when $\tau=2$ and $\alpha=0.1$ before and after rescaling to the worst case.}
\end{table}

We conclude from the tables the following points.
\begin{enumerate}
\item The method of \cite{Zhang} although provides clearly shorter confidence intervals for ranks, this comes at the cost of a low simultaneous coverage. Therefore, its results are generally unreliable and do not fulfill the requirement that it has a confidence level of at least $1-\alpha$.
\item The simultaneous coverage of the method of \cite{Zhang} increases as the range of means increases at a fixed $n$. On the other hand, it decreases as $n$ increases when the range of the means is held fixed.
\item The simultaneous coverage of our Tukey-based method increases with both the number of means $n$ and their range of values.
\item In average, our Tukey-based method seems to produce shorter CIs than the method of \cite{Zhang} when they are both rescaled, but this difference is not statistically significant. Indeed, the difference in efficiency was all the time within one simulation standard error. For example, when $\tau=0.5$ and $n=10$, we observed a standard error in $1-\hat{R}_n$ of about 0.01. We may state that when the two methods are properly rescaled, they do not have substantial differences.
\item Reducing the conservativeness of our Tukey-based method is always possible.
\item Repairing the anticonservativeness of the method of \cite{Zhang} is only possible in practice for $n\leq 50$.
\end{enumerate}
%%%%%%%%%%%%%%%%%%%%%%%%%%%%%%%%%%%%%%%%%%%%%%%%%%%%%%%%%%%%%%%%%%%%
%\subsection{Swedish hospitals}
%Do it for the data.
%Use the data to calculate coverage as if they were true means.
%Attractive but very dangerous and 

%%%%%%%%%%%%%%%%%%%%%%%%%%%%%%%%%%%%%%%%%%%%%%%%%%%%%%%%%%%%%%%%%%%%
\subsection{The case of different standard deviations: A dataset on hospitals in the Netherlands}
We study a dataset for Dutch hospitals concerning abdominal aneurysms surgery. The study includes 9489 patients operated at 64 hospitals in the Netherlands at dates mostly between the years 2012 and 2016. The number of patients per hospital ranged from 3 to 358 with an average of 150 patients per hospital. The dataset includes the following variables
\begin{itemize}
\item the hospital ID where the patient was treated;
\item the date of surgery;
\item the context of surgery: Elective, Urgent, Emergency;
\item the surgical procedure: "Endovascular", "Endovascular converted" and "Open". "Endovascular" means the patient had a minimal invasive procedure through the femoral artery in the groin. "Endovasculair converted" means the surgeons first tried a minimal invasive procedure through the femoral artery in the groin, but then realized they had to do an open surgery;
\item a complication within 30 days (yes or no);
\item the mortality within 30 days (yes or no);
\item VpPOSSUM: a numerical score that summarizes the pre-operative state of the patient.
\end{itemize}
In order to conform to the normality assumption in our model, we exclude hospitals with small number of patients. This leaves us with 61 hospitals and each one of them has at least 54 patients. We compare these hospitals according to the mortality rate within 30 days. We correct for case-mix effect with a fixed effect logistic regression model using the \texttt{VpPOSSUM} variable. One of the hospitals has no patients who died within 30 days after surgery. Thus, we added to all the hospitals a row of data with a virtual patient who died within 30 days after surgery and with a value of \texttt{VpPOSSUM} equals to the average in the corresponding hospital. This prevents the logistic regression from getting an infinite standard error for this hospital. Besides, the influence on the other hospitals is rather minor because of the relatively high number of patients in them. 

Before we apply the methods we presented in this paper, we calculate the rescaled significance level at the worst case configuration, that is we consider a null vector of means and the vector of standard deviations ordered according to the worst configuration we found in paragraph (\ref{subsec:RescalUneqSigma}), namely configuration (\ref{eqn:ConfigTree1}). We use the first 10 hospitals, 30 hospitals and finally all the hospitals and see how the rescaled significance level changes. 
\begin{table}[ht]
\centering
\begin{tabular}{cccccccccccc}
 & Tukey & Zhang \\
$n=10$ & 0.246 & 0.0003\\  
$n=30$ & 0.432 & $<5\times 10^{-6}$ \\
$n=61$ & 0.583 & $<5\times 10^{-6}$ 
\end{tabular}
\caption{Rescaled significance level $\tilde{\alpha}$ in order to get an actual simultaneous coverage equal to $90\%$ at the worst case configuration for the dataset of Dutch hospitals when considering the mortality as the primary outcome.}
\label{tab:RescaledAlphaDataMort}
\end{table}

\begin{figure}[ht]
\centering
\includegraphics[scale = 0.5]{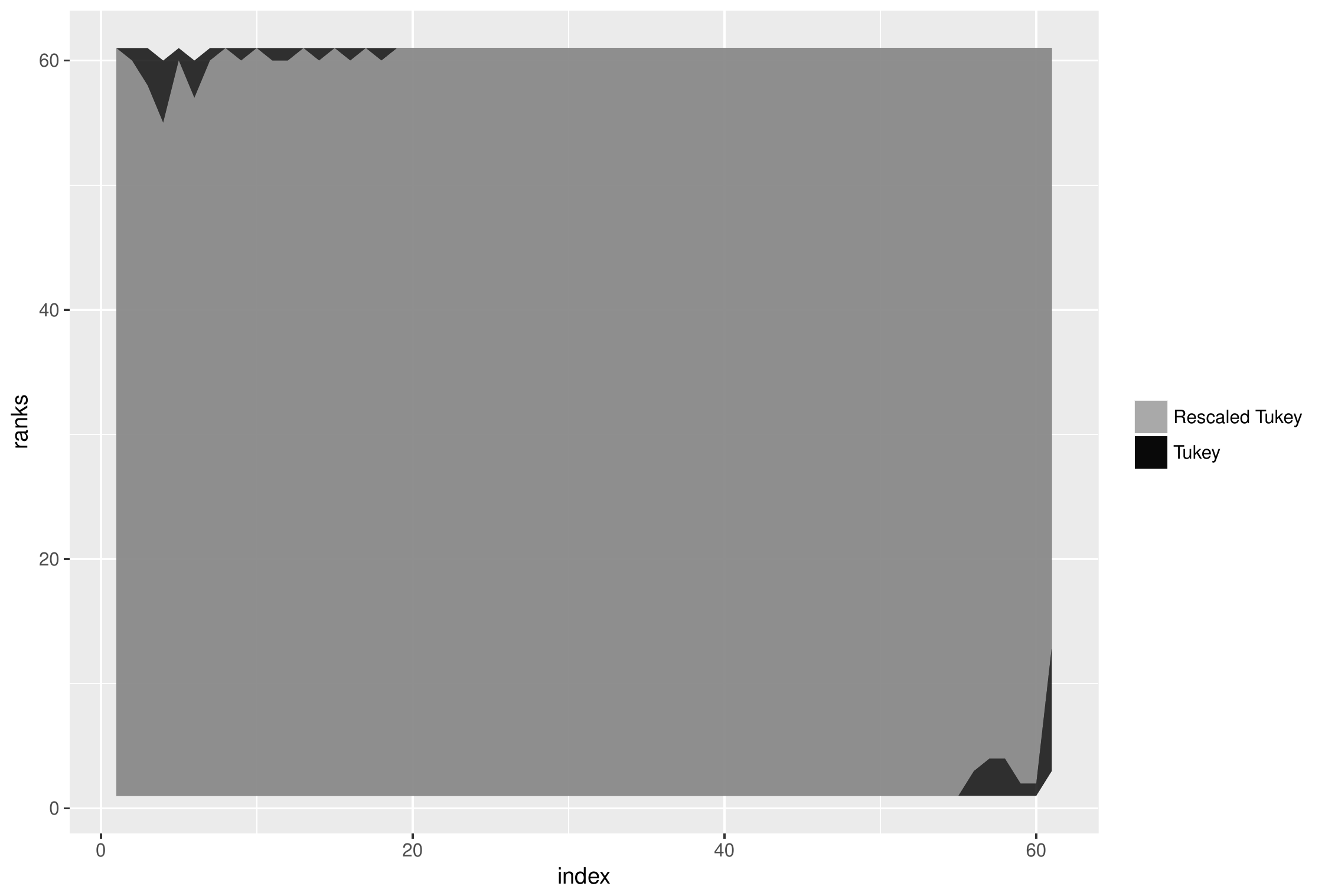}
\caption{Simultaneous confidence intervals for the ranks of 61 hospitals in the Netherlands with joint level $90\%$. The performance indicator is the mortality rate, and the hospital effect is corrected for case-mix using a logistic regression. The hospitals are not distinguishable using the mortality rate. Tukey and its sequential variant gave identical results.}
\label{fig:AdjustedDSSATukVsRescaledMortality}
\end{figure}

In order to apply the Monte-Carlo method of \cite{Zhang} and make sure that the confidence level is at least $90\%$, we need to use a significance level below $5\times 10^{-6}$ (table (\ref{tab:RescaledAlphaDataMort})). Using any value for $\tilde{\alpha}$ superior than $5\times 10^{-6}$ on the full data would result in a miss leading conclusion, because there is no guarantee that the simultaneous confidence level is actually at least $90\%$. Therefore, we avoid using it on the full dataset and only apply our Tukey-based methods. The simultaneous CIs for the ranks of the hospitals at the joint level $90\%$ are illustrated in figure (\ref{fig:AdjustedDSSATukVsRescaledMortality}). 

\begin{figure}[ht]
\centering
\includegraphics[scale = 0.5]{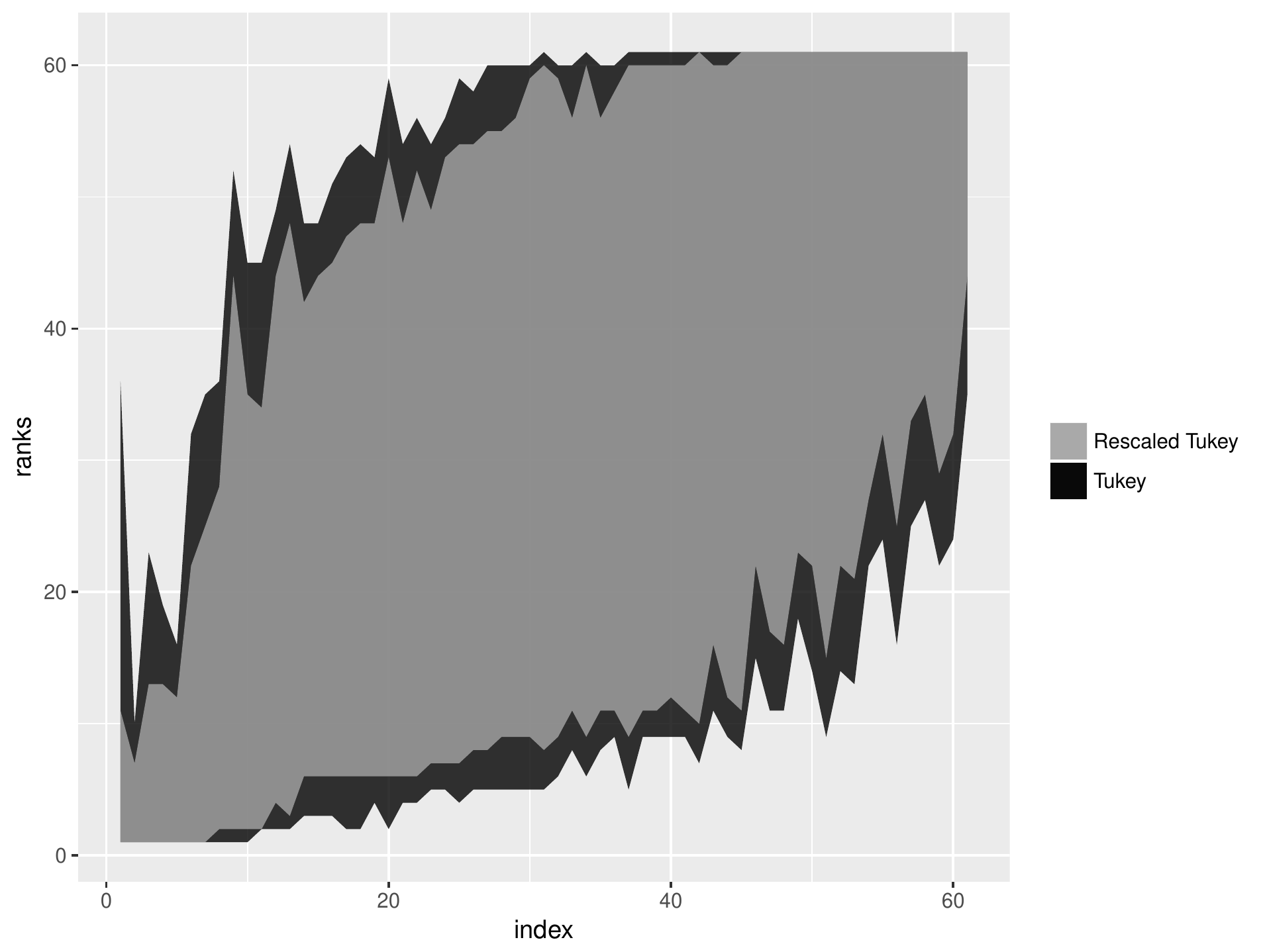}
\caption{Simultaneous confidence intervals for the ranks of 61 hospitals in the Netherlands. Data is corrected for case-mix effect.}
\label{fig:AdjustedDSSATukVsRescaledSurg}
\end{figure}

The confidence intervals cover the whole range of ranks, and there are barely any differences among the hospitals according to the mortality rate. The rankability is $0.001$ for the Tukey-based method without rescaling, and is $0.012$ after rescaling. This can either be normal, that is all Dutch hospitals have the same performance, or due to a low power of our methods. In order to find out, we change the output variable in the logistic regression model and correct for case-mix effects with the type of surgery as an output. The resulting CIs at joint level $90\%$ for the ranks are illustrated in figure (\ref{fig:AdjustedDSSATukVsRescaledSurg}) with a rankability of 0.240 for Tukey's HSD. Rescaling the significance level clearly improves the results of our Tukey-based CIs. The rescaled significance level is $\tilde{\alpha} = 0.607$. The new rankability is $0.358$. Here again, we could not apply the method of \cite{Zhang} because we obtain similar results to table (\ref{tab:RescaledAlphaDataMort}). We may state that for only 7 hospitals we find that they may get the first rank, and that for the remaining 54 hospitals we can confidently state they are not first rank.

\begin{figure*}[t!]
\centering
\begin{subfigure}{0.45\textwidth}
\includegraphics[scale = 0.5]{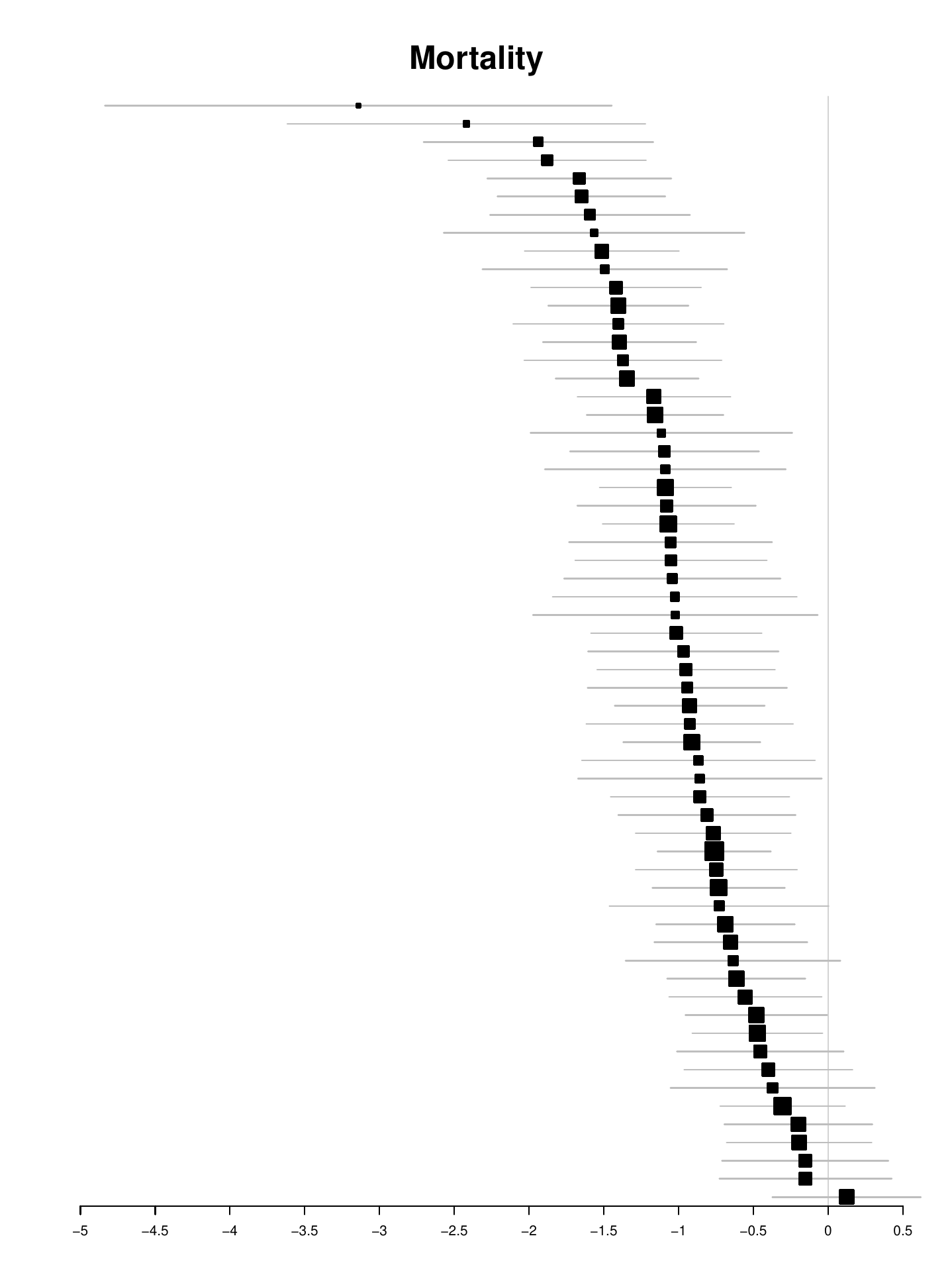}
%\caption{Forest plot for the hospital}
%\label{fig:ForestPlotMort}
\end{subfigure}
\qquad
\begin{subfigure}{0.45\textwidth}
\includegraphics[scale = 0.5]{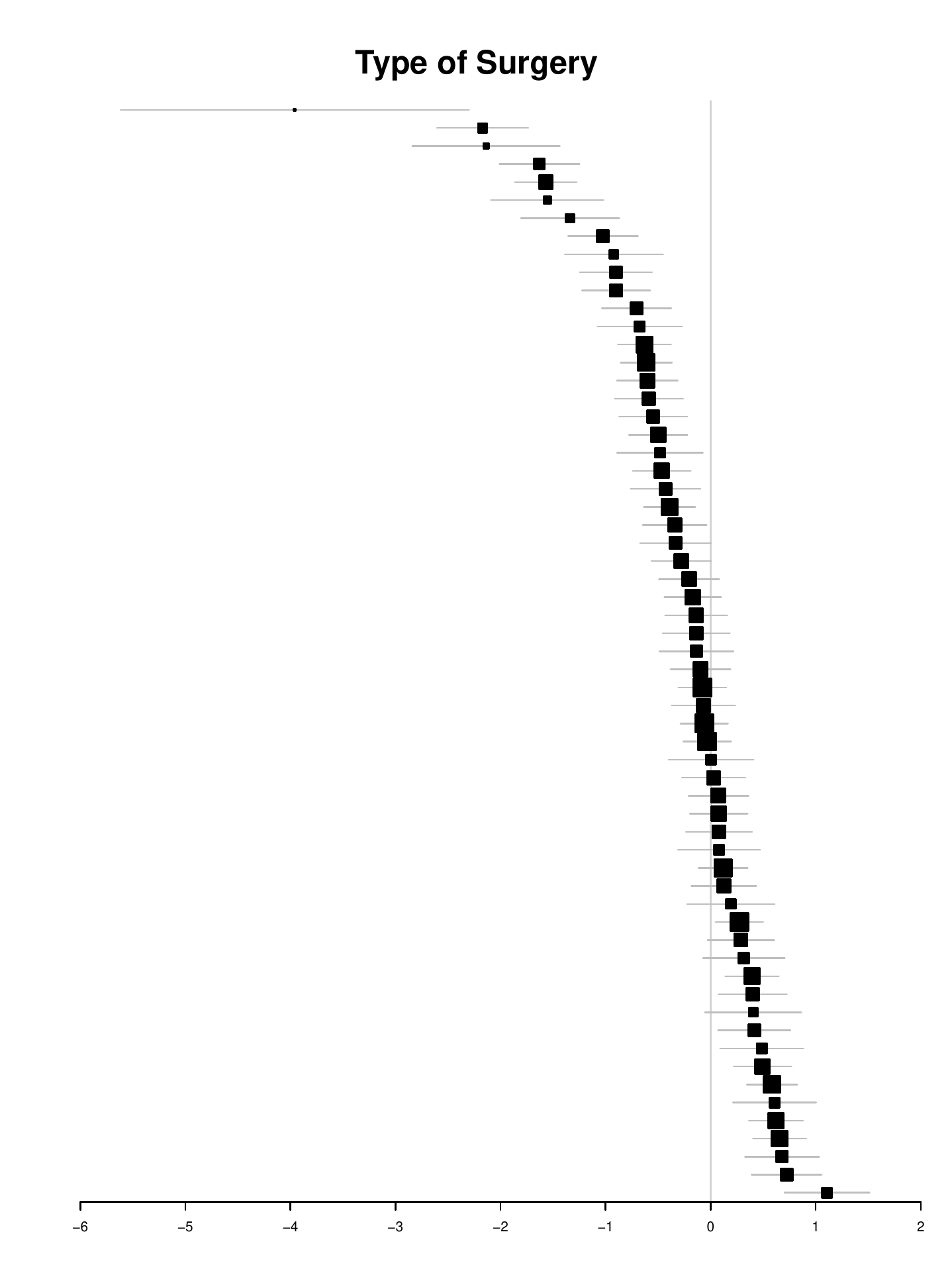}
%\label{fig:ForestPlotSurg}
\end{subfigure}
\caption{Forest plots for the hospitals effect after case-mix correction based on the mortality rate or the surgical procedure. The hospital effects based on the mortality rate are less distinguishable (have wider CIs) than the ones based on the surgical procedure.}
\label{fig:ForestPlotMortSurg}
\end{figure*}

We make a forest plot for the hospital effect after case-mix correction for both the mortality withing 30 days and the surgical procedure. Figure (\ref{fig:ForestPlotMortSurg}) shows that indeed the mortality rate induces very few differences among the hospitals whereas the type of surgery seems to show more differences. We also fit a random effect mixed-model to the hospital effect for both choices (separately) using function \texttt{rma} from package \texttt{metafor} (\citet{MetaforPackage}) and estimate the variance of the random effects using the Sidiki-Jonkman method and test for heterogeneity among the hospitals. We find a p-value of $0.951$ while using the mortality rate and a p-value lower than $0.001$ while using the type of surgery. This supports our claim that the reason behind the very wide CIs for ranks when we use the mortality rate is not the low power of our method but is rather that there are not much differences in the data. 

\section{Discussion}
We presented a novel method to produce simultaneous CIs for ranks based on Tukey's HSD and proposed a practical improvement under Assumption 1 that there are no ties among the means. The only method in the literature (as far as we know) claiming to produce simultaneous confidence intervals for ranks is the simulation-based method of \citet{Zhang}. We showed through simulations that the simultaneous confidence level goes below the nominal level unless the means are very far from each other. We proposed a solution to fix this problem by rescaling the confidence level. Surprisingly, after rescaling, the results of our Tukey-based method and the method of \cite{Zhang} seem to be almost the same and the differences are not significant.

By providing valid methods for simultaneous CIs for ranks, practitioners are provided with a way which looks at all the institutions together instead of only looking at a specific one. Simultaneity also provides a way to state which of the institutions could be ranked first best, second best, and so one, and which of these institutions may not get to be first rank. These pieces of information could not be obtained only be looking at pointwise CIs for the ranks. 

We found it impractical to rescale the significance level for \cite{Zhang}'s method when the number of means is higher than 30 whereas our Tukey-based method is always rescalable. While comparing the performance of 64 Dutch, it was only possible to use our Tukey-based method. The comparison of the hospitals according to their mortality rate at 30 days after surgery showed no differences among the hospitals. However, by considering the type of surgery as our primary output, some differences among the hospitals became quite clear (especially the extremities) and we were able for example to detect 7 hospitals which have the chance to be first rank. 

Using Dunnet's test, it is possible to look for a confidence interval for the rank of only one prespecified institution that we are interested in. On the other hand, in our approach we considered two-sided CIs. We could also be interested in looking only at worse or better institutions and thus gain more power by only considering one-sided CIs.

\bigskip
\begin{center}
{\large\bf SUPPLEMENTARY MATERIAL}
\end{center}
\section{Proofs of Propositions}
\subsection{Proof of proposition \ref{prop:TukeySimultCIs}}\label{Append:TukeySimultCIs}
%\begin{proof}
From Tukey's procedure, we can obtain simultaneous confidence intervals for the differences between the centers at level $1-\alpha$, that is $\mu_i-\mu_j$ for $i,j\in\{1,\cdots,n\}$, see \citet[sec. 2.1]{HochbergBook}. In other words, we have
\begin{equation}
\mathbb{P}\left(\mu_i - \mu_j \in \left[y_i - y_j \pm \sqrt{\sigma_i^2 + \sigma_j^2}q_{1-\alpha}\right], \forall i,j\right)\geq 1-\alpha.
\label{eqn:SimCIsDiff}
\end{equation}
Denote $[a_{i,j},b_{i,j}]$ the confidence interval for the difference $\mu_i-\mu_j$ in the previous display. Define also $L_i = 1+\#\{j:\; a_{i,j}>0\}$ and $U_i=n-\#\{j:\; b_{i,j}\leq 0\}$.
%In order to draw a simultaneous confidence statement about the ranks, we use the implication
%\[\left.\begin{array}{c} \mu_i - \mu_1 \in [a_{i,1},b_{i,1}] \\
%\vdots \\
%\mu_{i} - \mu_n \in [a_{i,n},b_{i,n}] \end{array} \right\} \Rightarrow l_i \geq L_i, u_i\leq U_i,\]
%where $l_i$ and $u_i$ are the lower and upper ranks (\ref{eqn:LowerRank},\ref{eqn:UpperRank}) of center $\mu_i$, and $L_i = 1+\#\{j:\; a_{i,j}>0\}$ and $U_i=n-\#\{j:\; b_{i,j}\leq 0\}$.
Let $E_i = \{\mu_i - \mu_j\in[a_{i,j},b_{i,j}], \forall j\neq i\}$. It is easy to see that the event $E_i$ implies the event $\{l_i\geq L_i, u_i\leq U_i\}$ for any $i$. Thus using inequality (\ref{eqn:SimCIsDiff}), we may write 
\[\mathbb{P}\left(\forall i, l_i\geq L_i, u_i\leq U_i\right) \geq \mathbb{P}\left(\forall i\neq j, \mu_i - \mu_j\in[a_{i,j},b_{i,j}]\right) \geq 1-\alpha.\]
Hence, the confidence intervals for the set-ranks $[L_i,U_i]$ have a joint level of at least $1-\alpha$. \\
%\end{proof}

%%%%%%%%%%%%%%%%%%%%%%%%%%%%%%%%%%%%%%%%%%%%%%
% ...................................
%%%%%%%%%%%%%%%%%%%%%%%%%%%%%%%%%%%%%%%%%%%%%%

\subsection{Proof of proposition \ref{prop:TukeyRankExactCov}}
%\begin{proof}
Under the full null, the set-rank of any $\mu_i$ is $[1,n]$. We need to show that the event
\begin{equation}
\left\{L_i\leq 1, U_i\geq n, \forall i\right\}
\label{eqn:EventRank}
\end{equation}
implies the event
\begin{equation}
\mu_i - \mu_j \in \left[y_i - y_j \pm \sqrt{\sigma_i^2 + \sigma_j^2}q_{1-\alpha}\right].
\label{eqn:EventDiff}
\end{equation}
Since $L_i\in\{1,\cdots,n\}$ and $U_i\in\{1,\cdots,n\}$, then the event (\ref{eqn:EventRank}) is equivalent to
\[\{L_i=1+\#\{j\neq i: y_i\geq y_j - \sqrt{\sigma_i^2 + \sigma_j^2}q_{1-\alpha}\}=1, U_i = n-\#\{j\neq i: y_i\leq y_j + \sqrt{\sigma_i^2 + \sigma_j^2}q_{1-\alpha}\} = n, \forall i\}.\]
The first part is equivalent to the event that there is no $j$ such that $\{y_i\geq y_j - \sqrt{\sigma_i^2 + \sigma_j^2}q_{1-\alpha}\}$ occurs. Similar reasoning for the second part entails that the event (\ref{eqn:EventRank}) is equivalent to
\[\{\forall j\neq i, y_i\leq y_j - \sqrt{\sigma_i^2 + \sigma_j^2}q_{1-\alpha}, y_i\geq y_j + \sqrt{\sigma_i^2 + \sigma_j^2}q_{1-\alpha}\}\] 
which is clearly the same as the event (\ref{eqn:EventDiff}).
%\end{proof}

%%%%%%%%%%%%%%%%%%%%%%%%%%%%%%%%%%%%%%%%%%%%%%
% ...................................
%%%%%%%%%%%%%%%%%%%%%%%%%%%%%%%%%%%%%%%%%%%%%%

\section{R Code for the Calculus of the Coverage of Zhang et al.'s Algorithm}\label{Append:SimCoverage}
\subsection{Calculating the coverage}
\begin{verbatim}
library(ICRanks)
n = 10; TrueCenters = 1:n
# Take a subset and generate the data
alpha = 0.05; sigma = rep(1,n)

K = 10^4
coverage = 100
coverageTuk = 100
for(i in 1:100)
{
y = as.numeric(sapply(1:n, function(ll) rnorm(1,TrueCenters[ll],sd=sigma[ll])))
ind = sort.int(y, index.return = T)$ix
y = y[ind]
resZhang = BootstrCIs(y, sigma, alpha = 0.05, N = K, K = K, maxiter = 10)
resTukey = ic.ranks(y, sigma, Method = "Tukey", alpha = 0.05)

if(sum(ind<resZhang$Lower | ind>resZhang$Upper)>0) 
	coverage = coverage - 1
if(sum(ind<resTukey$Lower | ind>resTukey$Upper)>0) 
	coverageTuk = coverageTuk - 1
}
\end{verbatim}

\subsection{An R function to calculate the CIs for the ranks according to Zhang et al.'s method}
\begin{verbatim}
BootstrCIs = function(y, sigma, alpha=0.05, N = 10^4, K = N, precision = 1e-6,
							maxiter = 50)
{
# A function which calculates the individual CIs at level beta
Spiegelhalter = function(mus,ses,beta, N = 10^4)
{
k=length(mus)
r=apply(x,2,rank)
r=apply(r,1,quantile,probs=c(beta/2,1-beta/2),type=3)
df=data.frame(lower=r[1,],upper=r[2,])
return(df)
}
n = length(y)
beta1 = 0; beta2 = alpha
beta = (beta2 + beta1) / 2
x=sigma*matrix(rnorm(K*n),nrow=n) + y
InitCIs = Spiegelhalter(y, sigma, alpha, N)
counter = 0; coverage = K
while(abs(beta1 - beta2)>precision | counter<=maxiter)
{

 # Generate individual CIs at level beta
 res = Spiegelhalter(y, sigma, beta, N)
 # Check the coverage
 coverage = K
 for(j in 1:K)
 {
	ind = rank(x[,j])
	if(sum(ind<res$lower | ind>res$upper)>0) coverage = coverage - 1
 }
 if(coverage/K >= 1-alpha)
 {
	beta1 = beta
 }else
 {
	beta2 = beta
 }
  beta = (beta2 + beta1) / 2

 counter = counter + 1
}
if(coverage/K < 1-alpha) beta = beta1
res = Spiegelhalter(y, sigma, beta, N)
return(list(Lower = res$lower, Upper = res$upper, coverage = coverage/K))
}

\end{verbatim}

%\begin{description}
%
%\item[Title:] Brief description. (file type)
%
%\item[R-package for  MYNEW routine:] R-package ÒMYNEWÓ containing code to perform the diagnostic methods described in the article. The package also contains all datasets used as examples in the article. (GNU zipped tar file)
%
%\item[HIV data set:] Data set used in the illustration of MYNEW method in Section~ 3.2. (.txt file)
%
%\end{description}

%\section{BibTeX}
%
%We hope you've chosen to use BibTeX!\ If you have, please feel free to use the package natbib with any bibliography style you're comfortable with. The .bst file agsm has been included here for your convenience. 

\clearpage
\bibliographystyle{plainnat}
\bibliography{biblioFile}

\end{document}